\documentclass[twocolumn,superscriptaddress,preprintnumbers,amsmath,amssymb,prb]{revtex4}

\usepackage{graphicx}
\usepackage{dcolumn}
\usepackage{color}
\usepackage{bm}
\usepackage{booktabs}

\newcommand{\upa}{\uparrow}
\newcommand{\dna}{\downarrow}

\newcommand{\beeq}{\begin{eqnarray}}
\newcommand{\eneq}{\end{eqnarray}}

\begin{document}


\title{Hidden charge-conjugation, parity, and time-reversal symmetries and massive Goldstone (Higgs) modes in superconductors}

\author{Shunji Tsuchiya}
\email{tsuchiya@phys.chuo-u.ac.jp}
\affiliation{Department of Physics, Chuo University, 1-13-27 Kasuga, Bunkyo-ku, Tokyo 112-8551, Japan}
\affiliation{Research and Education Center for Natural Sciences, Keio University, Hiyoshi 4-1-1, Yokohama, Kanagawa 223-8521, Japan}

\author{Daisuke Yamamoto}
\affiliation{Department of Physics and Mathematics, Aoyama Gakuin University, 5-10-1 Fuchinobe, Chuo-ku, Sagamihara, Kanagawa 252-5258, Japan}

\author{Ryosuke Yoshii}
\affiliation{Department of Physics, Chuo University, 1-13-27 Kasuga, Bunkyo-ku, Tokyo 112-8551, Japan}
\affiliation{Research and Education Center for Natural Sciences, Keio University, Hiyoshi 4-1-1, Yokohama, Kanagawa 223-8521, Japan}

\author{Muneto Nitta}
\affiliation{Department of Physics, Keio University, Hiyoshi 4-1-1, Yokohama, Kanagawa 223-8521, Japan}
\affiliation{Research and Education Center for Natural Sciences, Keio University, Hiyoshi 4-1-1, Yokohama, Kanagawa 223-8521, Japan}

\date{\today}

\begin{abstract}
A massive Goldstone (MG) mode, often referred to as a Higgs amplitude mode, is a collective excitation that arises in a system involving spontaneous breaking of a continuous symmetry, along with a gapless Nambu-Goldstone mode. It has been known in the previous studies that a pure amplitude MG mode emerges in superconductors if the dispersion of fermions exhibits the particle-hole (p-h) symmetry. However, clear understanding of the relation between the symmetry of the Hamiltonian and the MG modes has not been reached. Here we reveal the fundamental connection between the discrete symmetry of the Hamiltonian and the emergence of pure amplitude MG modes. To this end, we introduce nontrivial charge-conjugation ($\mathcal C$), parity ($\mathcal P$), and time-reversal ($\mathcal T$) operations that involve the swapping of pairs of wave vectors symmetrical with respect to the Fermi surface. The product of $\mathcal{CPT}$ (or its permutations) represents an exact symmetry analogous to the CPT theorem in the relativistic field theory. It is shown that a fermionic Hamiltonian with a p-h symmetric dispersion exhibits the discrete symmetries under $\mathcal C$, $\mathcal P$, $\mathcal T$, and $\mathcal{CPT}$. We find that in the superconducting ground state, $\mathcal T$ and $\mathcal P$ are spontaneously broken simultaneously with the U(1) symmetry. Moreover, we rigorously show that amplitude and phase fluctuations of the gap function are uncoupled due to the unbroken $\mathcal C$. In the normal phase, the MG and NG modes become degenerate, and they have opposite parity under $\mathcal T$. Therefore, we conclude that the lifting of the degeneracy in the superconducting phase and the resulting emergence of the pure amplitude MG mode can be identified as a consequence of the the spontaneous breaking of $\mathcal T$ symmetry but not of $\mathcal P$ or U(1).
\end{abstract}
\maketitle

\section{Introduction} 
Massive Goldstone (MG) modes, often referred to as Higgs amplitude modes, and Nambu-Goldstone (NG) modes are ubiquitous in systems that involve spontaneous breaking of continuous symmetries \cite{goldstone-61,goldstone-62,higgs-64,pekker-15}. In the simplest U(1) symmetry breaking, the former induce amplitude oscillation of a complex order parameter \cite{higgsdef} and the latter induce phase oscillation.
Whereas NG modes have been studied in various condensed matter systems, MG modes have evaded observations until recently with only a few exceptions \cite{sooryakumar-81,giannetta-80,demsar-99}.
\par
Despite the increasing number of observations, for example, in superconductors \cite{sooryakumar-81,matsunaga-13,matsunaga-14,measson-14,sherman-15,katsumi-18}, quantum spin systems \cite{ruegg-08,jain-17,hong-17,souliou-17}, charge-density-wave materials \cite{demsar-99,yusupov-10}, and ultracold atomic gases \cite{bissbort-11,endres-12,leonard-17,behrle-18}, and theoretical studies \cite{littlewood-81,engelbrecht-97,tsuchiya-13,gazit-13,cea-15,nakayama-15,bjerlin-16, krull-16,moor-17,liberto-18}, fundamental aspects of MG modes in condensed matter systems have not been fully understood, in contrast to NG modes; spontaneous breaking of a continuous symmetry does not guarantee emergence of MG modes, while that of NG modes is ensured by the Goldstone theorem \cite{goldstone-62}. 
For instance, whereas a MG mode appears in a Bardeen-Cooper-Schrieffer (BCS) superconductor \cite{sooryakumar-81,littlewood-81}, it does not exist in a Bose-Einstein condensate (BEC) \cite{varma-02}, despite the fact that both of the systems involve U(1) symmetry breaking and furthermore one evolves continuously to the other through the BCS-BEC crossover \cite{leggett-80,nozieres-85,sademelo-93,ohashi-02,regal-04}. Varma pointed out that the approximate particle-hole (p-h) symmetry, i.e., the linearly approximated fermionic dispersion $\xi_{\bm k}\simeq v_F(k-k_F)$ ($v_F$ is the Fermi velocity and $k_F$ is the Fermi wave number), results in the effective Lorentz invariance of the time-dependent Ginzburg-Landau equation in the weak-coupling BCS limit, which yields the decoupled amplitude and phase modes \cite{varma-02}. A pure amplitude MG mode also appears in lattice systems if the energy bands exhibit the rigorous p-h symmetry \cite{tsuchiya-13,cea-15}.
\par
It has been thus recognized in the previous studies that a pure amplitude MG mode emerges in superconductors if the dispersion of fermions $\xi_{\bm k}$ exhibits the p-h symmetry \cite{littlewood-81,engelbrecht-97,varma-02,tsuchiya-13,cea-15}. However, the p-h symmetry in the context of the previous works refers to the characteristic feature of the fermionic dispersion $\xi_{\bm k}$ that should be distinguished from the symmetry of the Hamiltonian. Meanwhile, clear understanding of the relation between {\it the symmetry of the Hamiltonian and MG modes} has not been reached.
\par
In this paper, we reveal the fundamental connection between {\it the discrete symmetry of the Hamiltonian and the emergence of pure amplitude MG modes}. We introduce three discrete operations for general non-relativistic systems of fermions, which we refer to ``charge-conjugation" ($\mathcal C$), ``parity" ($\mathcal P$), and ``time-reversal" ($\mathcal T$). The product of $\mathcal{CPT}$ (or its permutations) represents an exact symmetry analogous to the CPT theorem in the relativistic field theory \cite{lee-81}. We show that the standard BCS Hamiltonian with a p-h symmetric dispersion is invariant under $\mathcal C$, $\mathcal P$, $\mathcal T$, and $\mathcal{CPT}$ in addition to the global U(1) gauge invariance. If the U(1) symmetry is spontaneously broken in the superconducting ground state, the symmetries under $\mathcal P$ and $\mathcal T$ are simultaneously broken while the symmetry under $\mathcal C$ is unbroken. We rigorously show that amplitude and phase fluctuations of the gap function are uncoupled due to the unbroken $\mathcal C$. The MG mode thus induces pure amplitude oscillations of the gap function in a p-h symmetric system. It is also shown that the MG and NG modes reduce to the degenerate states in the normal phase due to the U(1) symmetry and they have opposite parity under $\mathcal T$. Therefore, the lifting of the degeneracy in the superconducting phase and the resulting emergence of the pure amplitude MG mode can be identified as a consequence of the the spontaneous breaking of $\mathcal T$ symmetry but not of $\mathcal P$ or U(1). Thus, the breaking of $\mathcal T$ proves to be responsible for the emergence of the pure amplitude MG mode.
\par
This paper is organized as follows: In Sec.~II, we present the model and introduce the pseudospin representation. In Sec.~III, we define the three discrete operations $\mathcal C$, $\mathcal T$, and $\mathcal P$ to discuss the symmetries of the Hamiltonian under the operations of $\mathcal C$, $\mathcal T$, $\mathcal P$, and $\mathcal{CPT}$. In Sec.~IV, we study the symmetry of the superconducting ground state. In Sec.~V, we discuss collective modes within the classical spin analysis. In Sec.~VI, we give a rigorous proof of the uncoupled amplitude and phase fluctuations of the gap function in a p-h symmetric system due to the unbroken $\mathcal C$. In Sec.~VII, we give a direct demonstration of the relation between the emergence of the pure amplitude MG mode and the spontaneously broken $\mathcal T$ symmetry. In Sec.~VIII, we summarize.  We set $\hbar=k_{\rm B}=1$ throughout the paper.
\section{Pseudospin representation}
We study for simplicity the reduced BCS Hamiltonian \cite{fullH}
\begin{eqnarray}
\mathcal H&=&\sum_{\bm k,s}\xi_{\bm k}c_{\bm k s}^\dagger
 c_{\bm k s} -g \sum_{\bm k,\bm k'} c_{\bm k\upa}^\dagger c_{-\bm
 k\dna}^\dagger c_{-\bm k'\dna}c_{\bm k'\upa},
\label{eq.HBCS}
\end{eqnarray}
where $c_{\bm k s}^\dagger$ ($c_{\bm k s}$) is the creation (annihilation) operator of a fermion with momentum $\bm k$ and spin $s$ $(=\uparrow,\downarrow)$, $g(>0)$ denotes the attractive interaction between fermions, and $\xi_{\bm k}=\varepsilon_{\bm k}-\mu$ is the kinetic energy of a fermion measured from the chemical potential $\mu$. For example, $\varepsilon_{\bm k}=k^2/2m$ in a continuous system ($m$ is the mass of a fermion). We do not specify the form of $\varepsilon_{\bm k}$ for generality of argument.
\par
To discuss the symmetries of the Hamiltonian (\ref{eq.HBCS}), it is convenient to introduce the pseudospin representation \cite{anderson-58}: $S_{\mu\bm k}=\frac{1}{2}\Psi_{\bm k}^\dagger\tau_{\mu}\Psi_{\bm k}$ ($\mu=x,y,z$),
where $\bm\tau=(\tau_x,\tau_y,\tau_z)$ are Pauli matrices and $\Psi_{\bm k}=(c_{\bm k\upa},c_{-\bm k\dna}^\dagger)^t$ is the Nambu spinor \cite{nambu-60}.
Note that $S_{z\bm k}$ is related to the fermion number operator $n_{\bm k s}=c_{\bm k s}^\dagger c_{\bm k s}$ by $S_{z\bm k}=\frac{1}{2}(n_{\bm k\upa}+n_{\bm k\dna}-1)$. In the pseudospin language, the fermion vacuum is the spin-down state  ($|0\rangle_{\bm k}=|\!\dna\rangle_{\bm k}$) and the fully occupied state is the spin-up state  ($c_{\bm k\upa}^\dagger c_{-\bm k\dna}^\dagger|0\rangle_{\bm k}=|\!\upa\rangle_{\bm k}$). 
\par
The pseudospin representation of the Hamiltonian (\ref{eq.HBCS}) is given by
\begin{eqnarray}
\mathcal H=\sum_{\bm k}2\xi_{\bm k} S_{z\bm k}-g\sum_{\bm k,\bm k'}\bm S_{\perp\bm k}\cdot \bm S_{\perp\bm k'},
\label{eq.HBCS_spin}
\end{eqnarray}
where $\bm S_{\perp\bm k}=(S_{x\bm k},S_{y\bm k})$. The kinetic energy (interaction) term is translated into the Zeeman (ferromagnetic XY exchange) term in the pseudospin language. The rotational symmetry of the Hamiltonian (\ref{eq.HBCS_spin}) in the $xy$-plane represents the U(1) symmetry of Eq.~(\ref{eq.HBCS}) with respect to the  transformation $\Psi_{\bm k}\to e^{i\tau_z\alpha}\Psi_{\bm k}$.

\section{Hidden discrete symmetries}

In this section, we define three discrete transformations for fermions and discuss the symmetry of the Hamiltonian (\ref{eq.HBCS_spin}) under those operations.

\subsection{Charge-conjugation}

\begin{figure}
\centering
\includegraphics[width=6cm]{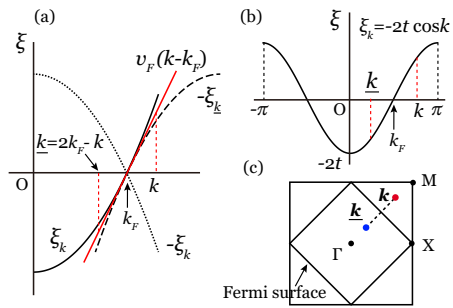}
\caption{Illustration of the wave vector $\underline{\bm k}$ and the dispersion $-\xi_{\underline{\bm k}}$  in (a) a continuous system and (b) the 1D lattice at half-filling ($\mu=0$). (c) $\underline{\bm k}$ for the half-filled energy band in the square lattice.} 
\label{fig.phsymmetry}
\end{figure}

Let us consider a unitary transformation for the Nambu spinor \cite{serene-83}:
\begin{eqnarray}
{\mathcal C} \Psi_{\bm k} {\mathcal C}=\tau_x\Psi_{\underline{\bm k}},\quad {\mathcal C} \Psi_{\bm k}^\dagger {\mathcal C}=\Psi_{\underline{\bm k}}^\dagger\tau_x .
\end{eqnarray}
Here, $\underline{\bm k}$ is the mirror reflected wave vector of $\bm k$ with respect to the Fermi surface, i.e., $\bm k$ and $\underline{\bm k}$ are on the opposite side of the Fermi surface and away from it with the same minimum distance (see Figs.~\ref{fig.phsymmetry} (a)-(c)). For example, $\underline{\bm k} =(2k_F-k)\bm k/|\bm k|$ in a continuous system. Note that $\underline{\bm k}=\bm k$ if $\bm k$ is on the Fermi surface.
Since $\mathcal C$ transforms a particle ($c^\dagger$) into a hole ($c$) and {\it vice versa}, it can be referred to as a ``charge conjugation" operation. 
$\mathcal C$ is specifically given by
\begin{eqnarray}
\mathcal C={\mathcal F}\prod_{\bm k}\sigma_{x\bm k},\quad \mathcal F=\prod_{\xi_{\bm k}>0} f_{\bm k,\underline{\bm k}},\label{eq.operatorC}
\end{eqnarray}
where $\sigma_{\mu\bm k}=2S_{\mu\bm k}$. The operator $f_{\bm k,\underline{\bm k}}$ swaps the state of $\bm k$ and that of $\underline{\bm k}$: $f_{\bm k,\underline{\bm k}}|\psi\rangle_{\bm k}|\phi\rangle_{\underline{\bm k}}= |\phi\rangle_{\bm k}|\psi\rangle_{\underline{\bm k}}$. One can show $\mathcal C^\dagger=\mathcal C$ and ${\mathcal C}^2=1$ from  Eq.~(\ref{eq.operatorC}).
\par
The pseudospin operators are transformed by $\mathcal C$ as
\begin{eqnarray}
{\mathcal C}S_{\mu\bm k}{\mathcal C}=(-1)^{\delta_{\mu,x}+1}S_{\mu\underline{\bm k}},\quad{\mathcal C}S_{\mu}{\mathcal C}=(-1)^{\delta_{\mu,x}+1}S_{\mu},\label{eq.CSkC}
\end{eqnarray}
where $S_\mu=\sum_{\bm k}S_{\mu\bm k}$ is the total spin.
Equation~(\ref{eq.CSkC}) shows that $\mathcal C$ consists of the $\pi$ rotation of pseudospins about the $x$-axis and the swapping of $\bm k$ and $\underline{\bm k}$.
\par
Transforming Eq.~(\ref{eq.HBCS_spin}) by $\mathcal C$, we obtain
\begin{eqnarray}
{\mathcal C}\mathcal H{\mathcal C}=\sum_{\bm k}2(-\xi_{\underline {\bm k}})S_{z\bm k}-g\sum_{\bm k,\bm k'}{\bm S}_{\perp\bm k}\cdot{\bm S}_{\perp\bm k'}.\label{eq.CHC}
\end{eqnarray}
Hence, ${\mathcal C}{\mathcal H}{\mathcal C}={\mathcal H}$ and equivalently $[{\mathcal H},\mathcal C]=0$, if the fermion dispersion satisfies the condition
\begin{equation}
-\xi_{\underline{\bm k}}=\xi_{\bm k}.\label{eq.phcondition}
\end{equation}
Equation~(\ref{eq.phcondition}) indicates the invariance of the dispersion $\xi_{\bm k}$ under the successive mirror reflections with respect to $\xi=0$ and $k=k_F$ (see Figs.~\ref{fig.phsymmetry} (a) and \ref{fig.phsymmetry} (b)), which we refer to {\it particle-hole (p-h) symmetry} in view of the fact that the density of states $N(\xi)=\sum_{\bm k}\delta(\xi-\xi_{\bm k})$ is even if Eq.~(\ref{eq.phcondition}) holds. 
\par
Figure~\ref{fig.phsymmetry}(a) shows that, whereas $\xi_{\bm k}=k^2/2m-\mu$ is not p-h symmetric, the linearized dispersion $\xi_{\bm k}\simeq v_F(k-k_F)$ is p-h symmetric. Therefore, a continuous system has an approximate p-h symmetry if the interaction is weak enough. On the other hand, Fig.~\ref{fig.phsymmetry} (b) illustrates that the tight-binding energy band in the $d$-dimensional cubic lattice $\xi_{\bm k}=-2t\sum_{i=1}^d\cos(k_i)$ ($t$ is the hopping matrix element) exhibits a rigorous p-h symmetry at half-filling ($\mu=0$).

\subsection{Time-reversal}
The ``time-reversal" operation of the Nambu spinor and the pseudospin operators are defined to be
\begin{eqnarray}
&&{\mathcal T}\Psi_{\bm k}{\mathcal T}^{-1}=\tau_y\Psi_{\underline{\bm k}},\quad {\mathcal T}\Psi_{\bm k}^\dagger {\mathcal T}^{-1}=\Psi_{\underline{\bm k}}^\dagger\tau_y,\\
&&{\mathcal T}S_{\mu\bm k} {\mathcal T}^{-1}=-S_{\mu\underline{\bm k}},\quad \mathcal T{S_\mu}{\mathcal T}^{-1}=-S_\mu.\label{eq.TST}
\end{eqnarray}
The time-reversal $\mathcal T$ can be written in the form
\begin{eqnarray}
\mathcal T =U_T\mathcal K,\quad U_T={\mathcal F}\prod_{\bm k}(-i\sigma_{y\bm k}),\label{eq.operatorT}
\end{eqnarray}
where $\mathcal K$ is the complex conjugation operator and $U_T$ is the unitary operator that rotates pseudospins $\pi$ about the $y$-axis and swaps $\bm k$ and $\underline{\bm k}$. 
From Eq.~(\ref{eq.TST}), the p-h symmetric Hamiltonian that satisfies Eq.~(\ref{eq.phcondition}) is time-reversal invariant ${\mathcal T}\mathcal H{\mathcal T}^{-1}
=\mathcal H$. $\mathcal T$ reverses the time in the Heisenberg representation as ${\mathcal T}S_{\mu}(t){\mathcal T}^{-1}=-S_{\mu}(-t)$.
\par
It is important to note that $\mathcal T$ represents ``time-reversal" in the pseudospin space, which is different from the usual time-reversal operation discussed, for example, in Ref.~\onlinecite{sigrist-91}. Although the usual time-reversal symmetry is not broken in $s$-wave superconductors \cite{sigrist-91}, $\mathcal T$ is spontaneously broken simultaneously with the U(1) symmetry breaking as we shall show later.

\subsection{Parity} 
The ``parity" operation, denoted by $\mathcal P$, is defined to be the inversion of pseudospins in the $xy$-plane. It is equivalent to the $\pi$ rotation about the $z$-axis and therefore can be represented as
\begin{equation}
\mathcal P=\prod_{\bm k}\sigma_{z\bm k}.\label{eq.operatorP}
\end{equation}
It satisfies $\mathcal P^\dagger=\mathcal P$ and ${\mathcal P}^2=1$. The transformation by $\mathcal P$ is given as
\begin{eqnarray}
\mathcal P\Psi_{\bm k}{\mathcal P}=\tau_z\Psi_{\bm k}, \quad {\mathcal P}\Psi_{\bm k}^\dagger {\mathcal P}=\Psi_{\bm k}^\dagger\tau_z,\\
{\mathcal P}S_{\mu\bm k}{\mathcal P}=(-1)^{\delta_{\mu,z}+1}S_{\mu\bm k},\  {\mathcal P}S_{\mu}{\mathcal P}=(-1)^{\delta_{\mu,z}+1}S_{\mu}.
\end{eqnarray}
The Hamiltonian (\ref{eq.HBCS_spin}) is invariant by $\mathcal P$: $\mathcal P\mathcal H\mathcal P={\mathcal H}$. Since the $\pi$ rotation in the $xy$-plane is an element of U(1), $\mathcal P$ is trivially broken in the U(1) broken ground state. 

\subsection{CPT invariance}
The transformation by the product $\Theta={\mathcal C}{\mathcal P}{\mathcal T}$ is given as
\begin{eqnarray}
\Theta\Psi_{\bm k}\Theta^{-1}=i\Psi_{\bm k}, \quad \Theta\Psi_{\bm k}^\dagger\Theta^{-1}=-i\Psi_{\bm k}^\dagger,\\
{\Theta}S_{\mu\bm k}{\Theta}=(-1)^{\delta_{\mu,y}+1}S_{\mu\bm k},\ {\Theta}S_{\mu}{\Theta}=(-1)^{\delta_{\mu,y}+1}S_{\mu}.
\end{eqnarray}
Using Eqs.~(\ref{eq.operatorC}), (\ref{eq.operatorT}), and (\ref{eq.operatorP}), we obtain $\Theta=\prod_{\bm k}(-1)\cdot\mathcal K$ and thus $\Theta \mathcal H\Theta^{-1}=\mathcal H$.
Since the Lagrangian 
$\mathcal L=\sum_{\bm k}i\Psi_{\bm k}^\dagger\frac{\partial }{\partial t}\Psi_{\bm k}-{\mathcal H}$
is transformed as
$\Theta\mathcal L(t)\Theta^{-1}=\mathcal L(-t)$, the action is invariant and therefore $\mathcal C\mathcal P\mathcal T$ and all other permutations of $\mathcal C$, $\mathcal P$, and $\mathcal T$ are exact symmetries analogous to the CPT invariance in relativistic systems \cite{lee-81}.

\section{Symmetry of the ground state}

\begin{figure}
\centering
\includegraphics[width=8.5cm]{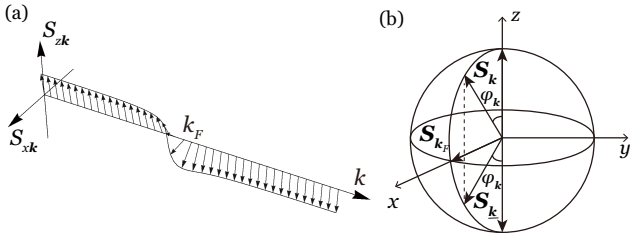}
\caption{Schematic illustration of the pseudospin distribution $\bm S_{\bm k}$ described by the BCS wave function $|\Psi\rangle$ for a positive real gap function (a) as a function of $k$ \cite{anderson-58} and (b) on the Bloch sphere. (a) Spins rotate in the $xz$-plane from up to down towards the positive $x$-direction as $k$ increases from below to above $k_F$. (b) In a p-h symmetric system, ${\bm S}_{\underline{\bm k}}$ is the mirror reflected image of $\bm S_{\bm k}$ with respect to the $xy$-plane.}
\label{fig.spinconfig}
\end{figure}

We study the symmetries of the superconducting ground state focusing on that of a p-h symmetric system. It is reasonable to expect that all the symmetries of the true ground state are realized in the BCS wave function $|\Psi\rangle=\prod_{\bm k}(u_{\bm k}|\dna\rangle_{\bm k}+v_{\bm k}|\upa\rangle_{\bm k})$. Here, $u_{\bm k}=\sqrt{(1+\xi_{\bm k}/E_{\bm k})/2}$ and $v_{\bm k}=\sqrt{(1-\xi_{\bm k}/E_{\bm k})/2}$. The gap function is set positive real in the ground state without loss of generality $\Delta_0=g\sum_{\bm k}\langle c_{-\bm k\dna} c_{\bm k\upa} \rangle=g\langle S_{x}\rangle>0$. $E_{\bm k}=\sqrt{\xi_{\bm k}^2+\Delta_0^2}$ is the dispersion of single-particle excitations (bogolons). 
$|\Psi\rangle$ represents the ground state of the mean-field (MF) Hamiltonian ${\mathcal H}_{\rm MF}=-\sum_{\bm k}\bm H_{\bm k}^0\cdot\bm S_{\bm k}$, where ${\bm S}_{\bm k}=(S_{x\bm k},S_{y\bm k}, S_{z\bm k})$.
The effective magnetic field $\bm H_{\bm k}^0=(2\Delta_0,0,-2\xi_{\bm k})$ lies in the $xz$-plane with the polar angle $\varphi_{\bm k}$ (see Fig.~\ref{fig.spinconfig} (b)), where $\sin\varphi_{\bm k}=\Delta_0/E_{\bm k}$ and $\cos\varphi_{\bm k}=-\xi_{\bm k}/E_{\bm k}$. Note that $\varphi_{\underline{\bm k}}=\pi-\varphi_{\bm k}$, if Eq.~(\ref{eq.phcondition}) holds.
The requirement that the average spin $\bm S_{\bm k}^0=\langle \bm S_{\bm k}\rangle$ is in parallel with $\bm H_{\bm k}^0$ leads to the MF gap equation
$1=g\sum_{\bm k}\frac{1}{2E_{\bm k}}$ \cite{anderson-58}.
\par
Figures~\ref{fig.spinconfig}(a) and \ref{fig.spinconfig}(b) show the pseudospin configuration of the superconducting ground state described by $|\Psi\rangle$. The pseudospins smoothly rotate sidewise in the $xz$-plane  from up to down towards the positive $x$-direction as $k$ increases \cite{anderson-58}. The spontaneous U(1) symmetry breaking with respect to the phase of the gap function sets the direction of rotating spins projected in the $xy$-plane. In a p-h symmetric system, ${\bm S}_{\underline{\bm k}}$ is the mirror reflected image of $\bm S_{\bm k}$ with respect to the $xy$-plane. 
\par
The symmetry under $\mathcal C$ is unbroken in the ground state of a p-h symmetric system. In fact, using $u_{\underline{\bm k}}=v_{\bm k}$ and $v_{\underline{\bm k}}=u_{\bm k}$, the BCS wave function is shown to be parity even ($\mathcal C|\Psi\rangle=|\Psi\rangle$) that reflects the invariance of the MF Hamiltonian ($\mathcal C\mathcal H_{\rm MF}\mathcal C=\mathcal H_{\rm MF}$). As shown in Fig.~\ref{fig.spinconfig}, the pseudospin configuration is indeed invariant under the $\pi$ rotation of spins about the $x$-axis followed by the swapping of $\bm k$ and $\underline{\bm k}$.
In contrast, the symmetries under $\mathcal T$ and $\mathcal P$ are spontaneously broken accompanied with the U(1) symmetry breaking. The operation of either $\mathcal T$ or $\mathcal P$ flips the sign of the gap as
\begin{eqnarray}
{\mathcal T}{\mathcal H}_{\rm MF}{\mathcal T}^{-1}={\mathcal P}{\mathcal H_{\rm MF}}{\mathcal P}=-\sum_{\bm k}\bar{\bm H}^{0}_{\bm k}\cdot\bm S_{\bm k}=\bar{\mathcal H}_{\rm MF},\\
\mathcal T|\Psi\rangle=|\bar{\Psi}\rangle,\quad\mathcal P|\Psi\rangle=\left\{\prod_{\bm k}(-1)\right\}\cdot|\bar\Psi\rangle.
\end{eqnarray}
\begin{figure}
\centering
\includegraphics[width=5cm]{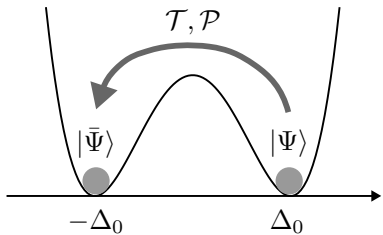}
\caption{Schematic illustration of the double-well potential for a real gap function and the spontaneous breaking of the symmetries under $\mathcal T$ and $\mathcal P$. The operation of either $\mathcal T$ or $\mathcal P$ flips the sign of the gap function and transforms $|\Psi\rangle$ to $|\bar\Psi\rangle$.}
\label{fig.doublewell}
\end{figure}
Figure~\ref{fig.doublewell} schematically illustrates the spontaneous breaking of the symmetry under $\mathcal T$ and $\mathcal P$ and their operations on $|\Psi\rangle$. Hereafter, the overline represents the replacement $\Delta_0\to -\Delta_0$, e.g., $\bar{\bm H}^{0}_{\bm k}=(-2\Delta_0,0,-2\xi_{\bm k})$ and $|\bar\Psi\rangle=\prod_{\bm k} (u_{\bm k}|\downarrow\rangle_{\bm k}-v_{\bm k}|\uparrow\rangle_{\bm k})$.
\par
The symmetries of the Hamiltonian and the ground state are compared between p-h symmetric and non-symmetric systems in Table~1. It shows that the broken symmetry of $\mathcal T$ and unbroken symmetry of $\mathcal C$ are characteristic to a p-h symmetric system. Given the fact that a pure amplitude MG mode arises only in a p-h symmetric system as shown later, Table~1 implies that it results from the broken $\mathcal T$ and $\mathcal C$, which we reveal in the following.
\begin{table}[tb]
\begin{tabular}{|c||cc|cc|} \hline
 & \multicolumn{2}{c|}{p-h symmetric} & \multicolumn{2}{c|}{p-h non-symmetric} \\ \cline{2-5}
Symmetry & $\mathcal H$ & $|\Psi\rangle$ & $\mathcal H$ & $|\Psi\rangle$ \\ \hline
$\mathcal C$ & \checkmark & \checkmark & $\times$ & $\times$ \\
$\mathcal T$ & \checkmark & $\times$ & $\times$ & $\times$ \\\hline
$\mathcal P$ & \checkmark & $\times$ & \checkmark & $\times$ \\
$\Theta=\mathcal{CPT}$ & \checkmark& \checkmark & \checkmark & \checkmark \\
U(1) & \checkmark &  $\times$ & \checkmark & $\times$\\ \hline
\end{tabular}
\caption{Symmetry of the Hamiltonian $\mathcal H$ and the ground state wave function $|\Psi\rangle$ for a p-h symmetric system ($\xi_{\bm k}=-\xi_{\underline{\bm k}}$) and a p-h non-symmetric system ($\xi_{\bm k}\neq-\xi_{\underline{\bm k}}$). \checkmark and $\times$ mean presence and absence of the symmetry, respectively.}
\end{table}

\section{Collective modes}

We first discuss collective modes within the classical spin analysis~\cite{anderson-58} (Details are given in Appendix B). We study dynamics of the pseudospins based on the MF Hamiltonian
$\mathcal H_{\rm MF}'=-\sum_{\bm k}{\bm H}_{\bm k}\cdot{\bm S}_{\bm k}$.
Here, the magnetic field $\bm H_{\bm k}=(2{\rm Re}\Delta,-2{\rm Im}\Delta,-2\xi_{\bm k})$ is self-consistently determined by the gap function $\Delta=g\sum_{\bm k}\langle c_{-\bm k\downarrow}c_{\bm k\uparrow}\rangle=g(\langle S_x\rangle-i\langle S_y\rangle)$, which is allowed to take complex values. The time evolution of $\bm S_{\bm k}(t)$, which is treated as a classical spin, is described by the equation of motion
\begin{equation}
\frac{d\bm S_{\bm k}}{dt}={\bm S}_{\bm k}\times\bm H_{\bm k}.\label{eq.Skt}
\end{equation}
\par
Introducing amplitude and phase fluctuations from the ground state $\Delta=(\Delta_0+\delta \Delta)e^{i\delta\theta}$, one finds that spin fluctuations in the $x$-direction induce amplitude fluctuations $\delta \Delta =g\delta S_x$ and those in the $y$-direction induce phase fluctuations $\delta\theta=-g\delta S_y/\Delta_0$, where $\delta\bm S_{\bm k}=\bm S_{\bm k}(t)-{\bm S}_{\bm k}^0$.
Linearizing Eq.~(\ref{eq.Skt}) by fluctuations $\delta\Delta,\delta\theta\propto e^{-i\omega t}$, we obtain
\begin{eqnarray}
\left(1-2g\chi_{xx}(\omega)\right)\delta \Delta-2g\chi_{xy}(\omega)\Delta_0\delta\theta=0,\label{eq.deltaSx}\\
2g\chi_{yx}(\omega)\delta\Delta-\left(1-2g\chi_{yy}(\omega)\right)\Delta_0\delta\theta=0,\label{eq.deltaSy}
\end{eqnarray}
where $\chi_{\mu\nu}(\omega)$ are the dynamical spin susceptibilities defined as
$\chi_{\mu\nu}(\omega)=-i\int_0^\infty \langle[S_\nu,S_\mu(t)]\rangle e^{-i\omega t}dt$ ($S_\mu(t)$ is the Heisenberg representation and $\langle\dots\rangle$ denotes the average). For example, $\chi_{xy}$ represents the coupling of amplitude and phase, while $\chi_{zx}$ represents that of density and amplitude. The susceptibilities are calculated as
\begin{eqnarray}
&&\chi_{xx}=\sum_{\bm k}\frac{\xi_{\bm k}^2}{E_{\bm k}(4E_{\bm k}^2-\omega^2)},\!\chi_{yy}=\sum_{\bm k}\frac{E_{\bm k}}{4E_{\bm k}^2-\omega^2},\label{eq.suscept_xxyy}\\
&&\chi_{xy}=-\chi_{yx}=\frac{i\omega}{2}\sum_{\bm k}\frac{\xi_{\bm k}}{E_{\bm k}(4E_{\bm k}^2-\omega^2)}.\label{eq.suscept_xy}
\end{eqnarray} 
\par
Using the MF gap equation, one finds that Eqs.~(\ref{eq.deltaSx}) and (\ref{eq.deltaSy}) have the NG mode solution ($\delta\theta\neq0$, $\delta\Delta=0$) with $\omega=0$. They also have a solution for a pure amplitude mode ($\delta\Delta\neq 0$ and $\delta \theta=0$) with $\omega=2\Delta_0$, if phase and amplitude are uncoupled $\chi_{xy}(2\Delta_0)=\chi_{yx}(2\Delta_0)=0$. From Eq.~(\ref{eq.suscept_xy}), this leads to the condition
\begin{eqnarray}
\sum_{\bm k}\frac{1}{E_{\bm k}\xi_{\bm k}}=\int d\xi\frac{N(\xi)}{\xi\sqrt{\xi^2+\Delta_0^2}}=0.\label{eq.p-h}
\end{eqnarray}
Equation~(\ref{eq.p-h}) is satisfied if $N(\xi)$ is even. Thus, MG mode arises as a {\it pure amplitude mode} in a p-h symmetric system \cite{engelbrecht-97}. 
\par
The p-h symmetry also ensures fermion number conservation ($\delta S_z=0$) \cite{deltaSz0}. $\delta S_z$ is represented as 
\begin{eqnarray}
\delta S_z=2\chi_{zx}(\omega)\delta\Delta+2\chi_{zy}(\omega)\Delta_0\delta\theta,\label{eq.deltaSz}
\end{eqnarray}
where $\chi$s are given by
\begin{eqnarray}
\chi_{zx}=\sum_{\bm k}\frac{\Delta_0\xi_{\bm k}}{E_{\bm k}(4E_{\bm k}^2-\omega^2)},\!
\chi_{zy}=\sum_{\bm k}\frac{i\omega \Delta_0/2}{E_{\bm k}(4E_{\bm k}^2-\omega^2)}\label{eq.suscept_zxy}.
\end{eqnarray} 
The MG mode solution ($\delta\Delta\neq 0$, $\delta\theta=0$, and $\omega=2\Delta_0$) satisfies $\delta S_z=0$, if $\chi_{zx}(2\Delta_0)=0$, which reduces to Eq.~(\ref{eq.p-h}). Hence, the MG mode does not induce density fluctuation  and indeed conserves total fermion number $N=2S_z+\sum_{\bm k}1$.
\par
If the p-h symmetry is absent, due to $\chi_{zx}(\omega)\neq 0$ and $\chi_{zy}(\omega)\neq 0$, Eq.~(\ref{eq.deltaSz}) indicates that $\delta \Delta$ and $\delta \theta$ must be finite in order to satisfy $\delta S_z=0$. As a result, $\delta\Delta$ is inevitably coupled with $\delta \theta$ and therefore the MG mode induces both amplitude and phase fluctuations. The energy of the MG mode becomes greater than $2\Delta_0$ \cite{tsuchiya-13,cea-15}.

\section{Rigorous proof of $\chi_{xy}=\chi_{zx}=0$}

The arguments in the last section are based on the MF approximation restricted to zero temperature ($T=0$). We rigorously show that amplitude is decoupled from phase and density in a p-h symmetric system at any temperature. 
We focus on $\chi_{zx}(\omega)$ and evaluate $\langle [S_x,S_z(t)] \rangle\propto\sum_n e^{-E_n/T}\langle n|[S_x,S_z(t)]|n\rangle$. Here, $|n\rangle$ denotes an exact eigenstate of $\mathcal H$ with energy $E_n$. Since $\mathcal C$ is not broken, $|n\rangle$ is parity either even or odd under $\mathcal C$. Using the fact that $S_x$ and $S_z$ have opposite parity under $\mathcal C$, we obtain
\begin{eqnarray}
\langle n|S_xS_z(t)|n\rangle&=&(\langle n|{\mathcal C})({\mathcal C}S_x{\mathcal C})({\mathcal C}S_z(t){\mathcal C})({\mathcal C}|n\rangle)\nonumber\\
&=&-\langle n|S_xS_z(t)|n\rangle=0.
\end{eqnarray}
One can analogously show $\langle n|S_z(t)S_x| n\rangle=\langle n|S_zS_x(t)|n\rangle=\langle n|S_x(t)S_z|n\rangle=0$ and therefore $\chi_{zx}(\omega)=\chi_{xz}(\omega)=0$. 
$\chi_{xy}(\omega)=\chi_{yx}(\omega)=0$ can be shown analogously using the opposite parity of $S_x$ and $S_y$. Thus, the unbroken symmetry under $\mathcal C$ is essential for the pure amplitude character of the MG mode.

\section{Emergence of the MG mode by the broken $\mathcal T$ symmetry}

We show that the spontaneous breaking of $\mathcal T$ is responsible for the emergence of the MG mode.
The creation operator of the MG mode $\beta_{\rm H}^\dagger$ and that of the NG mode $\beta_{\rm NG}^\dagger$ derived by the Holstein-Primakoff theory are given by (see Appendix C for details) 
\begin{eqnarray}
&&\beta_{\rm H}^\dagger=A\sum_{\bm k}\frac{\xi_{\bm k}}{E_{\bm k}}\left(\frac{S'^+_{\bm k}}{2|\Delta_0|-2E_{\bm k}}+\frac{S'^-_{\bm k}}{2|\Delta_0|+2E_{\bm k}}\right),\label{eq.betaH}\\
&&\beta_{\rm NG}^\dagger=A'\sum_{\bm k}\frac{1}{E_{\bm k}}(S'^+_{\bm k}+S'^-_{\bm k}).\label{eq.betaNG}
\end{eqnarray}
Here, $S'^{\pm}_{\bm k}=S'_{x\bm k}\pm iS'_{y\bm k}$, which creates and annihilates a pair of bogolons, are the raising and lowering operators of the pseudospins for bogolons $\bm S'_{\bm k}=(S'_{x\bm k},S'_{y\bm k},S'_{z\bm k})$. $S_{\bm k}'^{\pm}$ are transformed as (see Appendix A)
\begin{eqnarray}
{\mathcal C}S_{\bm k}'^{\pm}{\mathcal C}=-S_{\underline{\bm k}}'^{\pm},\mathcal PS_{\bm k}'^\pm\mathcal P=- \bar{S}_{\bm k}'^\pm,\mathcal TS_{\bm k}'^\pm\mathcal T^{-1}=\bar{S}_{\underline{\bm k}}'^{\pm}.\label{eq.TSpT}
\end{eqnarray}
Using Eq.~(\ref{eq.TSpT}), one can show that the MG mode is even and the NG mode is odd under $\mathcal C$: 
\begin{equation}
{\mathcal C}\beta_{\rm H}^\dagger{\mathcal C}=\beta_{\rm H}^\dagger,\quad{\mathcal C}\beta_{\rm NG}^\dagger {\mathcal C}=-\beta_{\rm NG}^\dagger.\label{eq.CbetaC}
\end{equation} 
Their opposite parity under $\mathcal C$ is consistent with the uncoupled phase and amplitude. A single MG mode is thus prohibited to decay into odd number of NG modes by the selection rule. Moreover, since the excited states of energy $2\Delta_0$ with a pair of bogolons are odd under $\mathcal C$ (see Appendix A), a MG mode with energy $2\Delta_0$ is stable against decay into independent bogolons.
\par
The MG and NG modes thus have definite parity under $\mathcal C$ due to the unbroken $\mathcal C$, while the discrete symmetries under $\mathcal T$ and $\mathcal P$ are broken. From Eq.~(\ref{eq.TSpT}), we obtain 
\begin{eqnarray}
&&{\mathcal T}\beta_{\rm H}^\dagger {\mathcal T}^{-1}={\mathcal P}\beta_{\rm H}^\dagger {\mathcal P}=-\bar\beta_{\rm H}^\dagger,\label{eq.TbetaHT}\\
&&{\mathcal T}\beta_{\rm NG}^\dagger {\mathcal T}^{-1}=\bar\beta_{\rm NG}^\dagger,\quad{\mathcal P}\beta_{\rm NG}^\dagger {\mathcal P}=-\bar\beta_{\rm NG}^\dagger\label{eq.TbetaNGT},
\end{eqnarray}
where $\beta_{\rm H}^\dagger\to\bar\beta_{\rm H}^\dagger$ and $\beta_{\rm NG}^\dagger\to\bar\beta_{\rm NG}^\dagger$ by the replacement $\Delta_0\to-\Delta_0$.  
Note that using Eqs.~(\ref{eq.CbetaC}), (\ref{eq.TbetaHT}), and (\ref{eq.TbetaNGT}), $\Theta\beta_{\rm H}^\dagger\Theta^{-1}=\beta_{\rm H}^\dagger$ and $\Theta\beta_{\rm NG}^\dagger\Theta^{-1}=\beta_{\rm NG}^\dagger$ are indeed satisfied.
\par
Denoting the vacuum state for $\beta_{\rm H}$ and $\beta_{\rm NG}$ ($\bar\beta_{\rm H}$ and $\bar\beta_{\rm NG}$) as $|{\rm vac}\rangle$ ($|\overline{\rm vac}\rangle$), we have the relation $\mathcal T|{\rm vac}\rangle=\mathcal P|{\rm vac}\rangle=|\overline{\rm vac}\rangle$, since either $\mathcal T$ or $\mathcal P$ flips the sign of the gap function \cite{Pvac}.
From Eqs.~(\ref{eq.CbetaC}), (\ref{eq.TbetaHT}) and (\ref{eq.TbetaNGT}), one obtains
\begin{eqnarray}
&&\mathcal C(\beta_{\rm H}^\dagger|{\rm vac}\rangle)=\beta_{\rm H}^\dagger|{\rm vac}\rangle,\label{eq.CbetaH}\\
&&\mathcal T(\beta_{\rm H}^\dagger|{\rm vac}\rangle)=\mathcal P(\beta_{\rm H}^\dagger|{\rm vac}\rangle)=-\bar\beta_{\rm H}^\dagger|\overline{\rm vac}\rangle,\label{eq.TbetaH}\\
&&\mathcal C(\beta_{\rm NG}^\dagger|{\rm vac}\rangle)=-\beta_{\rm NG}^\dagger|{\rm vac}\rangle,\label{eq.CbetaNG}\\
&&\mathcal T(\beta_{\rm NG}^\dagger|{\rm vac}\rangle)=\bar\beta_{\rm NG}^\dagger|\overline{\rm vac}\rangle,\label{eq.TbetaNG}\\
&&\mathcal P(\beta_{\rm NG}^\dagger|{\rm vac}\rangle)=-\bar\beta_{\rm NG}^\dagger|\overline{\rm vac}\rangle.\label{eq.PbetaNG}
\end{eqnarray}
\par
In the normal phase, setting $\Delta_0=0$, $\beta_{\rm H}^\dagger|\rm vac\rangle$ and $\bar\beta_{\rm H}^\dagger|\overline{\rm vac}\rangle$ trivially reduce to the same state $\beta_{\rm H0}^\dagger|\rm FS\rangle\equiv|\phi_{\rm H}\rangle$, while $\beta_{\rm NG}^\dagger|\rm vac\rangle$ and $\bar\beta_{\rm NG}^\dagger|\overline{\rm vac}\rangle$ reduce to $\beta_{\rm NG0}^\dagger|{\rm FS}\rangle\equiv|\phi_{\rm NG}\rangle$. 
Here, $|{\rm FS}\rangle$ denotes the vacuum in the normal phase. $\beta_{\rm H0}^\dagger$ and $\beta_{\rm NG0}^\dagger$ are given by
\begin{eqnarray}
\beta_{\rm H0}^\dagger\equiv\left.\beta_{\rm H}^\dagger\right|_{\Delta_0=0}\propto\sum_{\bm k}\frac{1}{\xi_{\bm k}}S_{y\bm k}\label{eq.betaH0},\\
\beta_{\rm NG0}^\dagger\equiv\left.\beta_{\rm NG}^\dagger\right|_{\Delta_0=0}\propto\sum_{\bm k}\frac{1}{\xi_{\bm k}}S_{x\bm k}\label{eq.betaNG0}.
\end{eqnarray}
Since $\beta_{\rm NG0}^\dagger$ can be transformed to $\beta_{\rm H0}^\dagger$ by the $\pi/2$ rotation about the $z$-axis in the pseudospin space, Eqs.~(\ref{eq.betaH0}) and (\ref{eq.betaNG0}) indicate that the $|\phi_{\rm H}\rangle$ and $|\phi_{\rm NG}\rangle$ states are degenerate in the normal phase before breaking the U(1) symmetry \cite{cooperon,degeneracy}.
Setting $\Delta_0=0$ in Eqs.~(\ref{eq.CbetaH}), (\ref{eq.TbetaH}), (\ref{eq.CbetaNG}), (\ref{eq.TbetaNG}), and (\ref{eq.PbetaNG}), we obtain \cite{CPTcommutation}
\begin{eqnarray}
\mathcal C|\phi_{\rm H}\rangle=|\phi_{\rm H}\rangle,\ \mathcal T|\phi_{\rm H}\rangle=\mathcal P|\phi_{\rm H}\rangle=-|\phi_{\rm H}\rangle,\\
\mathcal T|\phi_{\rm NG}\rangle=|\phi_{\rm NG}\rangle,\ \mathcal C|\phi_{\rm NG}\rangle=\mathcal P|\phi_{\rm NG}\rangle=-|\phi_{\rm NG}\rangle.
\end{eqnarray}
The above equations show that $|\phi_{\rm H}\rangle$ is odd and $|\phi_{\rm NG}\rangle$ is even under $\mathcal T$. On the other hand, both $|\phi_{\rm H}\rangle$ and $|\phi_{\rm NG}\rangle$ are odd under $\mathcal P$. From these facts, we can conclude that the lifting of the degeneracy of $|\phi_{\rm H}\rangle$ and $|\phi_{\rm NG}\rangle$ in the superconducting phase
should be induced by the spontaneous breaking of $\mathcal T$ symmetry, not by the breaking of $\mathcal P$ or U(1) symmetry. Consequently, the breaking of $\mathcal T$ proves to be responsible for the emergence of the pure amplitude MG mode.
The spontaneously induced magnetic field that breaks the $\mathcal T$ symmetry is given by $H_{x\bm k}^0=2\Delta_0$. Therefore, the energy splitting between the MG and NG modes should be of the order of $|H_{x\bm k}^0|=2\Delta_0$. This is consistent with the fact that the energy gap of the MG mode is $2\Delta_0$.

\section{Conclusions} 
Extending the previous understanding of the emergence of the MG mode in the presence of the p-h symmetric fermionic dispersion, we have revealed the fundamental connection between the emergence of the pure amplitude MG mode and the discrete symmetry of the Hamiltonian in superconductors, which has not been clarified in the previous works. We have shown that a non-relativistic Hamiltonian for fermions with a p-h symmetric dispersion exhibits nontrivial discrete symmetries under $\mathcal C$, $\mathcal P$,  $\mathcal T$, and $\mathcal{CPT}$. In the U(1) broken superconducting ground state of such a p-h symmetric system, $\mathcal T$ and $\mathcal P$ are spontaneously broken, while $\mathcal C$ is unbroken. We have shown that the spontaneous breaking of the discrete $\mathcal T$ symmetry leads to the emergence of the MG mode that induces pure amplitude oscillation of the gap function due to the unbroken $\mathcal C$. It may be possible to show a similar relation between the discrete symmetry of the Hamiltonian and the emergence of the MG modes in other non-relativistic systems, such as ultracold bosons in optical lattices \cite{endres-12,liberto-18} and quantum spin systems \cite{jain-17,hong-17}.

\section*{Acknowledgments}
ST is grateful to C. A. R. S\'a de Melo, T. Nikuni, and N. Tsuji for inspiring discussions. DY thanks the support of CREST, JST No. JPMJCR1673, and of JSPS Grant-in-Aid for Scientific Research (KAKENHI Grant No. 18K03525). The work of RY and MN is supported by the Ministry of Education, Culture, Sports, Science (MEXT)-Supported Program for the Strategic Research Foundation at Private Universities ``Topological Science'' (Grant No. S1511006). The work of MN is also supported in part by
JSPS Grant-in-Aid for Scientific Research (KAKENHI Grant No.~16H03984 and 18H01217), and by a Grant-in-Aid for Scientific Research on Innovative Areas ``Topological Materials Science'' (KAKENHI Grant No.~15H05855) from the MEXT of Japan.

\appendix

\section{Pseudospin representation for bogolons}

In this Appendix, we introduce a pseudospin representation for bogolons and examine the symmetries of the states involving excited bogolons. The pseudospins for bogolons $\bm S'_{\bm k}=(S'_{x\bm k},S'_{y\bm k},S'_{z\bm k})$ \cite{anderson-58} are defined as
\begin{eqnarray}
\left(
\begin{array}{ccc}
S_{z\bm k}'\\
S_{x\bm k}'\\
S_{y\bm k}'
\end{array}
\right)=
\left(
\begin{array}{ccc}
-\cos\varphi_{\bm k} & -\sin\varphi_{\bm k} & 0\\
\sin\varphi_{\bm k} & -\cos\varphi_{\bm k} & 0\\
0 & 0 & 1
\end{array}
\right)
\left(
\begin{array}{cc}
S_{z\bm k}\\
S_{x\bm k}\\
S_{y\bm k}
\end{array}
\right).\label{eq.Sprime}
\end{eqnarray}
Using Eq.~(\ref{eq.Sprime}), the MF Hamiltonian is represented as
\begin{equation}
\mathcal H_{\rm MF}=\sum_{\bm k}2E_{\bm k}S'_{z\bm k}.
\end{equation}
\par
Denoting the eigenstates of $S_{z\bm k}'$ as $|\uparrow'\rangle_{\bm k}$ and $|\downarrow'\rangle_{\bm k}$, they can be written as
\begin{eqnarray}
|\uparrow'\rangle_{\bm k}&=&u_{\bm k}|\uparrow\rangle_{\bm k}-v_{\bm k}|\downarrow\rangle_{\bm k},\\
|\!\downarrow'\rangle_{\bm k}&=&u_{\bm k}|\downarrow\rangle_{\bm k}+v_{\bm k}|\uparrow\rangle_{\bm k},
\end{eqnarray}
where $|\downarrow'\rangle_{\bm k}$ represents the vacuum of bogolons and $|\uparrow'\rangle_{\bm k}$ the excited state of energy $2E_{\bm k}$, in which a pair of bogolons are excited. Since $\bm S_{\bm k}$ is rotated about the angle $\pi-\varphi_{\bm k}$ in the $xz$-plane in Eq.~(\ref{eq.Sprime}), all the rotated pseudospins ${\bm S}_{\bm k}'$ are aligned downward in the $z$ direction in the ground state. In fact, the BCS wave function can be written as
\begin{equation}
|\Psi\rangle=\prod_{\bm k}|\downarrow'\rangle.
\end{equation}
\par
The raising and lowering operators, which creates and annihilates a pair of bogolons, are given by
\begin{eqnarray}
S'^{\pm}_{\bm k}&=&S'_{x\bm k}\pm i S'_{y\bm k}\nonumber\\
&=&\frac{\xi_{\bm k}}{E_{\bm k}}S_{x\bm k}\pm iS_{y\bm k}+\frac{\Delta_0}{E_{\bm k}}S_{z\bm k}.
\end{eqnarray}
Equation~(\ref{eq.TSpT}) can be derived from the above equation.
\par
If we denote the excited state with a single pair of bogolons as
\begin{eqnarray}
|e_{\bm k}\rangle=S'^+_{\bm k}|\Psi\rangle=|\uparrow'\rangle_{\bm k}\prod_{\bm k'\neq\bm k}|\downarrow'\rangle_{\bm k'},
\end{eqnarray}
both $|e_{\bm k}\rangle$ and $|e_{\underline{\bm k}}\rangle$ have excitation energy $2E_{\bm k}$ and degenerate in a p-h symmetric system.
Using Eq.~(\ref{eq.TSpT}), we obtain
\begin{equation}
\mathcal C|e_{\bm k}\rangle = -|e_{\underline{\bm k}}\rangle.\label{eq.Ce}
\end{equation}
From Eq.~(\ref{eq.Ce}), it can be easily shown that $|e_{\bm k}\rangle -|e_{\underline{\bm k}}\rangle$ is parity even, while $|e_{\bm k}\rangle+|e_{\underline{\bm k}}\rangle$ is parity odd under $\mathcal C$. The even parity states vanish at $\bm k=\underline{\bm k}=\bm k_F$ because of $|e_{\bm k}\rangle =|e_{\underline{\bm k}}\rangle$. It means that the lower edge of the single-particle continuum with energy $2\Delta_0$ consists of parity odd states.

\section{Classical spin analysis}
\begin{figure}
\centering
\includegraphics[width=3cm]{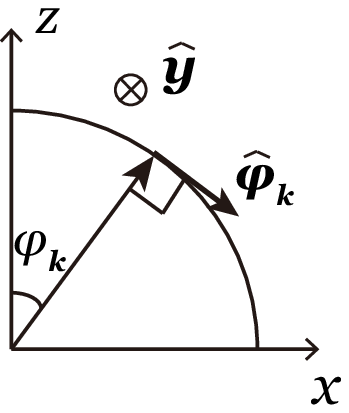}
\caption{Illustration of the unit vectors $\hat{{\bm \phi}}_{\bm k}$ and $\hat{\bm y}$. }
\label{fig.unitvector}
\end{figure}

In this Appendix, we give details of the classical spin analysis. Linearizing Eq.~(\ref{eq.Skt}) with respect to fluctuation $\delta\bm S_{\bm k}=\bm S_{\bm k}(t)-{\bm S}_{\bm k}^0$ ($\delta \bm H(t)=\bm H_{\bm k}(t)-\bm H_{\bm k}^0$), we obtain
\begin{eqnarray}
&&\frac{d}{dt}\delta S_{\parallel\bm k}=-\frac{1}{2}\delta H_y+H_{\bm k}^0\delta S_{y\bm k},\label{eq.Spara}\\
&&\frac{d}{dt}\delta S_{y\bm k}=\frac{1}{2}\delta H_x\cos\varphi_{\bm k}-H_{\bm k}^0\delta S_{\parallel\bm k}.\label{eq.Sy}
\end{eqnarray}
Here, we decompose the spin fluctuation into the two orthogonal directions as $\delta \bm S_{\bm k}=\delta S_{y\bm k}\hat{\bm y}+\delta S_{\parallel\bm k}\hat{\bm \varphi}_{\bm k}$. $\hat{\bm y}$ is the unit vector in the $y$-direction and $\hat {\bm \varphi}_{\bm k}=\cos\varphi_{\bm k}\hat{\bm x}-\sin\varphi_{\bm k}\hat{\bm z}$ is the unit vector illustrated in Fig.~\ref{fig.unitvector}. 
We note that
\begin{eqnarray}
&&\delta H_x=2g\delta S_x=2g\sum_{\bm k}\delta S_{\parallel\bm k}\cos\varphi_{\bm k},\label{deltaHx}\\
&&\delta H_y=2g\delta S_{y}=2g\sum_{\bm k}\delta S_{y\bm k}.\label{deltaHy}
\end{eqnarray}
\par
Using Eqs~(\ref{eq.Sy}) and (\ref{eq.Spara}), we obtain 
\begin{eqnarray}
&&\frac{d}{dt}\delta S_z=-\sum_{\bm k}\frac{d}{dt}\delta S_{\parallel\bm k}\sin\varphi_{\bm k}\nonumber\\
&&=\frac{1}{2}\delta H_y\sum_{\bm k}\sin\varphi_{\bm k}-2\Delta_0\sum_{\bm k}\delta S_{y\bm k}=0.\label{eq.deltaSz0}
\end{eqnarray}
Since $\delta S_z=0$ at the initial moment, Eq.~(\ref{eq.deltaSz0}) shows that the fermion number is conserved ($\delta S_z=0$) through the dynamics.
\par
If the gap function is constant in time, setting $\delta H_x=\delta H_y=0$ in Eqs.~(\ref{eq.Spara}) and (\ref{eq.Sy}), each pseudospin undergoes precession independently with frequency $\omega=2E_{\bm k}$. It represents a pair of bogolons arising from a broken Cooper pair.
\par
We consider collective dynamics of pseudospins involving nonzero $\delta H_x$ and/or $\delta H_y$. Assuming $\delta \bm S_{\bm k}(t), \delta \bm H(t)\propto e^{-i\omega t}$ in Eqs.~(\ref{eq.Sy}) and (\ref{eq.Spara}), we obtain 
\begin{eqnarray}
&&\delta S_{\parallel\bm k}=-\frac{\xi_{\bm k}}{4E_{\bm k}^2-\omega^2}\delta H_x+\frac{i\omega/2}{4E_{\bm k}^2-\omega^2}\delta H_y,\label{eq.Sparak}\\
&&\delta S_{y\bm k}=-\frac{i\omega\cos\varphi_{\bm k}/2}{4E_{\bm k}^2-\omega^2}\delta H_x+\frac{E_{\bm k}}{4E_{\bm k}^2-\omega^2}\delta H_y.\label{eq.Syk}
\end{eqnarray}
Substituting the above equations into Eqs.~(\ref{deltaHx}) and (\ref{deltaHy}), we obtain the coupled equations for $\delta H_x$ and $\delta H_y$ as
\begin{eqnarray}
&&\left(1-2g\chi_{xx}(\omega)\right)\delta H_x+2g\chi_{xy}(\omega)\delta H_y=0,\label{eq.deltaHx}\\
&&2g\chi_{yx}(\omega)\delta H_x+\left(1-2g\chi_{yy}(\omega)\right)\delta H_y=0,\label{eq.deltaHy}\\
&&\delta S_z=\chi_{zx}(\omega)\delta H_x-\chi_{zy}(\omega)\delta H_y, \label{eq.deltaHz}
\end{eqnarray}
where $\chi$s are given by Eqs.~(\ref{eq.suscept_xxyy}), (\ref{eq.suscept_xy}), and (\ref{eq.suscept_zxy}).
Equations~(\ref{eq.deltaSx}), (\ref{eq.deltaSy}), and (\ref{eq.deltaSz}) can be readily derived from Eqs.~(\ref{eq.deltaHx}), (\ref{eq.deltaHy}), and (\ref{eq.deltaHz}) by rewriting them in terms of $\delta\Delta$ and $\delta \theta$.
\par
\begin{figure}
\centering
\includegraphics[width=8cm]{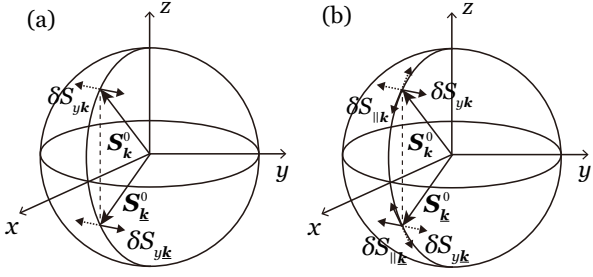}
\caption{Illustration of pseudospin oscillation induced by the NG mode (a) and the MG mode (b) in a system with p-h symmetric $\xi_{\bm k}$. (a) The NG mode induces in-phase oscillation of $\delta S_{y\bm k}$ and $\delta S_{y \underline{\bm k}}$. (b) The MG mode induces out-of-phase oscillation of $\delta S_{\parallel\bm k}$ and $\delta S_{\parallel\underline{\bm k}}$, as well as $\delta S_{y\bm k}$ and $\delta S_{y \underline{\bm k}}$. }
\label{fig.NGHiggsspin}
\end{figure}

If we set $\omega=0$ in Eqs.~(\ref{eq.deltaHx}) and (\ref{eq.deltaHy}), since $\chi_{xy}(0)=\chi_{yx}(0)=0$, $\delta H_x$ and $\delta H_y$ are uncoupled. Using $1-2g\chi_{yy}(0)= 0$ that reduces to the MF gap equation and $1-2g\chi_{xx}(0)\neq 0$, we obtain the solution for the NG mode: $\delta H_x=0$ and $\delta H_y\neq 0$ ($\delta\Delta=0$ and $\delta\theta\neq 0$). Equation~(\ref{eq.deltaHz}) indicates that the NG mode solution fulfills the number conservation $\delta S_z=0$, because $\chi_{zy}(0)=0$. From Eq.~(\ref{eq.Sparak}) and (\ref{eq.Syk}), one obtains
\begin{eqnarray}
\delta S_{\parallel\bm k}=0,\quad \delta S_{y\bm k}=\frac{1}{4E_{\bm k}}\delta H_y.
\end{eqnarray} 
The NG mode thus induces oscillations of pseudospins in the $y$-direction as illustrated in Fig.~\ref{fig.NGHiggsspin} (a). Since $\delta S_{y\underline{\bm k}}=\delta S_{y\bm k}$, the NG mode induces in-phase oscillation of $\delta S_{y\bm k}$ and $\delta S_{y\underline{\bm k}}$.
\par 
In a p-h symmetric system, if we set $\omega=2\Delta_0$ in Eqs.~(\ref{eq.deltaHx}) and (\ref{eq.deltaHy}), since $\chi_{xy}(2\Delta_0)=\chi_{yx}(2\Delta_0)=0$, $\delta H_x$ and $\delta H_y$ are uncoupled. Using $1-2g\chi_{xx}(2\Delta_0)= 0$ that reduces to the MF gap equation and $1-2g\chi_{yy}(0)\neq0$, we obtain the solution for the MG mode: $\delta H_x\neq 0$ and $\delta H_y= 0$ ($\delta\Delta\neq0$ and $\delta\theta= 0$). Equation~(\ref{eq.deltaHz}) indicates that the MG mode solution fulfills the number conservation $\delta S_z=0$, because $\chi_{zx}(2\Delta_0)=0$ if $\xi_{\bm k}$ satisfies Eq.~(\ref{eq.phcondition}). From Eq.~(\ref{eq.Sparak}) and (\ref{eq.Syk}), one obtains
\begin{eqnarray}
\delta S_{\parallel\bm k}=\frac{-1}{4\xi_{\bm k}}\delta H_x,\quad \delta S_{y\bm k}=\frac{i\Delta_0}{4E_{\bm k}\xi_{\bm k}}\delta H_x.
\end{eqnarray} 
The MG mode thus induces oscillations of pseudospins both in the $y$-direction and the direction of $\hat{\bm\phi}_{\bm k}$ as illustrated in Fig.~\ref{fig.NGHiggsspin} (b). Since $\delta S_{\parallel\underline{\bm k}}=-\delta S_{\parallel\bm k}$ and $\delta S_{y\underline{\bm k}}=-\delta S_{y\bm k}$, the MG mode induces out-of-phase oscillation of $\delta S_{\parallel\bm k}$ and $\delta S_{y\bm k}$. 

\section{Holstein-Primakoff theory}

In this Appendix, we develop the Holstein-Primakoff theory for the pseudospin Hamiltonian (\ref{eq.HBCS_spin}) to derive the creation and annihilation operators of the MG and NG modes. 
\par
Substituting Eq.~(\ref{eq.Sprime}) into Eq.~(\ref{eq.HBCS_spin}), one obtains
\begin{eqnarray}
&&\mathcal H=\sum_{\bm k}2\xi_{\bm k}(-\cos\varphi_{\bm k}S'_{z\bm k}+\sin\varphi_{\bm k}S'_{x\bm k})\nonumber\\
&&-g\sum_{\bm k,\bm k'}(\cos\varphi_{\bm k}\cos\varphi_{\bm k'}S'_{x\bm k}S'_{x\bm k'}+\sin\varphi_{\bm k}\cos\varphi_{\bm k'}\left\{S'_{z\bm k},S_{x\bm k'}\right\}\nonumber\\
&&+\sin\varphi_{\bm k}\sin\varphi_{\bm k'}S'_{z\bm k}S'_{z\bm k'}+S'_{y\bm k}S'_{y\bm k'}). \label{eq.HBCSdash}
\end{eqnarray}
Spin fluctuation can be quantized by the Holstein-Primakoff transformation \cite{holstein-40}:
\begin{eqnarray}
&&S'^+_{\bm k}=\alpha_{\bm k}^\dagger \sqrt{1-\alpha_{\bm k}^\dagger\alpha_{\bm k}},\label{eq.HPtrans1}\quad
S'^-_{\bm k}=(S'^+_{\bm k})^\dagger,\label{eq.HPtrans2}\\
&&S'_{z\bm k}=-\left(\frac{1}{2}-\alpha_{\bm k}^\dagger\alpha_{\bm k}\right),\label{eq.HPtrans3}
\end{eqnarray}
where $\alpha_{\bm k}^\dagger$ and $\alpha_{\bm k}$ denote, respectively, the creation and annihilation operators of a boson that represents spin fluctuation. They satisfy the usual commutation relations  $[\alpha_{\bm k},\alpha_{\bm k'}^\dagger]=\delta_{\bm k,\bm k'}$ and $[\alpha_{\bm k},\alpha_{\bm k'}]=[\alpha_{\bm k}^\dagger,\alpha_{\bm k'}^\dagger]=0$. When fluctuation is small $\alpha_{\bm k}^\dagger\alpha_{\bm k}\ll 1$, $S_{\bm k}'^+\simeq \alpha_{\bm k}^\dagger$ and $S_{\bm k}'^-\simeq \alpha_{\bm k}$ and therefore $\alpha_{\bm k}$ and $\alpha_{\bm k}^\dagger$ reduce to the annihilation and creation operators of a pair of bogolons, respectively.
\par
We expand Eq.~(\ref{eq.HBCSdash}) in terms of $\alpha_{\bm k}$ and $\alpha_{\bm k}^\dagger$. The zeroth and first order terms read
\begin{eqnarray}
\mathcal H_0&=&-\sum_{\bm k}\frac{\xi_{\bm k}^2}{E_{\bm k}}-\frac{\Delta_0^2}{g},\\
\mathcal H_1&=&\sum_{\bm k}(\xi_{\bm k}\sin\varphi_{\bm k}+\Delta_0\cos\varphi_{\bm k})(\alpha_{\bm k}+\alpha_{\bm k}^\dagger).
\end{eqnarray}
The first order term vanishes in the ground state $\mathcal H_1=0$ using $\sin\varphi_{\bm k}=\Delta_0/E_{\bm k}$ and $\cos\varphi_{\bm k}=-\xi_{\bm k}/E_{\bm k}$. 
The second order term reads
\begin{eqnarray}
&&\mathcal H_2=2\sum_{\bm k}E_{\bm k}\alpha_{\bm k}^\dagger\alpha_{\bm k}+\frac{g}{4}\sum_{\bm k,\bm k'}\left\{(1-\cos\varphi_{\bm k}\cos\varphi_{\bm k'})\right.\nonumber\\
&&\times\left.(\alpha_{\bm k}\alpha_{\bm k'}+\alpha_{\bm k}^\dagger\alpha_{\bm k'}^\dagger)-(1+\cos\varphi_{\bm k}\cos\varphi_{\bm k'})\right.\nonumber\\
&&\times\left.(\alpha_{\bm k}\alpha_{\bm k'}^\dagger+\alpha_{\bm k}^\dagger\alpha_{\bm k'})\right\}.\label{eq.H2}
\end{eqnarray}
We diagonalize $\mathcal H_2$ by a Bogoliubov transformation
\begin{eqnarray}
\beta_{\lambda}=\sum_{\bm k}(X_{\lambda\bm k}^*\alpha_{\bm k}+Y_{\lambda\bm k}^*\alpha_{\bm k}^\dagger),\label{eq.bosonbogoliubov1}\\
\beta_{\lambda}^\dagger=\sum_{\bm k}(X_{\lambda\bm k}\alpha_{\bm k}^\dagger+Y_{\lambda\bm k}\alpha_{\bm k}),\label{eq.bosonbogoliubov2}
\end{eqnarray}
where $\lambda$ labels the excited states. The bosonic operator $\beta_\lambda$ satisfies the commutation relations
\begin{eqnarray}
&&[\beta_\lambda,\beta_{\lambda'}^\dagger]=\sum_{\bm k}(X_{\lambda\bm k}^*X_{\lambda'\bm k}-Y_{\lambda\bm k}^*Y_{\lambda'\bm k})=\delta_{\lambda,\lambda'},\label{eq.normalcondition}\\
&&[\beta_\lambda^\dagger,\beta_{\lambda'}^\dagger]=\sum_{\bm k}(-X_{\lambda\bm k}Y_{\lambda'\bm k}+Y_{\lambda\bm k}X_{\lambda'\bm k})=0.
\end{eqnarray}
From Eqs.~(\ref{eq.bosonbogoliubov1})  and (\ref{eq.bosonbogoliubov2}), one can easily derive the inverse transformation
\begin{eqnarray}
\alpha_{\bm k}=\sum_{\lambda}(X_{\lambda\bm k}\beta_\lambda-Y_{\lambda\bm k}^*\beta_{\lambda}^\dagger),\\
\alpha_{\bm k}^\dagger=\sum_{\lambda}(X_{\lambda\bm k}^*\beta_\lambda^\dagger-Y_{\lambda\bm k}\beta_{\lambda}).
\end{eqnarray}
\par
Assuming that the second order term is diagonalized as $\mathcal H_2=\sum_\lambda\omega_\lambda\beta_\lambda^\dagger\beta_\lambda+{\rm const.}$, we obtain
\begin{eqnarray}
[\alpha_{\bm k},\mathcal H_2]=\sum_{\lambda}\omega_{\lambda}(X_{\lambda\bm k}\beta_{\lambda}+Y_{\lambda\bm k}^*\beta_{\lambda}^\dagger).
\label{eq.commute1}
\end{eqnarray}
On the other hand, using Eq.~(\ref{eq.H2}), one obtains
\begin{eqnarray}
[\alpha_{\bm k},\mathcal H_2]&=&\sum_{\lambda}\left\{\left(2E_{\bm k}X_{\lambda\bm k}-\frac{g}{2}\sum_{\bm k'}(\cos\varphi_{\bm k}\cos\varphi_{\bm k'}+1)X_{\lambda\bm k'}\right.\right.\nonumber\\
&&\left.+\frac{g}{2}\sum_{\bm k'}(\cos\varphi_{\bm k}\cos\varphi_{\bm k'}-1)Y_{\lambda\bm k'}\right)\beta_{\lambda}\nonumber\\
&&+\left(-2E_{\bm k}Y^*_{\lambda\bm k}-\frac{g}{2}\sum_{\bm k'}(\cos\varphi_{\bm k}\cos\varphi_{\bm k'}-1)X^*_{\lambda\bm k'}\right.\nonumber\\
&&\left.\left.+\frac{g}{2}\sum_{\bm k'}(\cos\varphi_{\bm k}\cos\varphi_{\bm k'}+1)Y^*_{\lambda\bm k'}\right)\beta^\dagger_{\lambda}\right\}.
\label{eq.commute2}
\end{eqnarray}
Comparing Eqs.~(\ref{eq.commute1}) and (\ref{eq.commute2}), one obtains sets of equations for $X_{\lambda\bm k}$ and $Y_{\lambda\bm k}$ as
\begin{eqnarray}
&&2E_{\bm k}X_{\lambda\bm k}-\frac{g}{2}\left\{(a_\lambda-c_\lambda)\cos\varphi_{\bm k}+(b_\lambda+d_\lambda)\right\}\nonumber\\
&&=\omega_{\lambda}X_{\lambda\bm k},\label{eq.equationX}\\
&&-2E_{\bm k}Y_{\lambda\bm k}-\frac{g}{2}\left\{(a_\lambda-c_\lambda)\cos\varphi_{\bm k}-(b_\lambda+d_\lambda)\right\}\nonumber\\
&&=\omega_{\lambda}Y_{\lambda\bm k},\label{eq.equationY}
\end{eqnarray}
where the coefficients $a_\lambda$, $b_\lambda$, $c_\lambda$, and $d_\lambda$ are given by
\begin{eqnarray}
a_\lambda=\sum_{\bm k}\cos\varphi_{\bm k}X_{\lambda \bm k},\quad
b_\lambda=\sum_{\bm k}X_{\lambda \bm k},\\
c_\lambda=\sum_{\bm k}\cos\varphi_{\bm k}Y_{\lambda \bm k},\quad
d_\lambda=\sum_{\bm k}Y_{\lambda \bm k}.
\end{eqnarray}
Equations~(\ref{eq.equationX}) and (\ref{eq.equationY}) can be formally solved as
\begin{eqnarray}
X_{\lambda\bm k}=\frac{g}{2}\frac{(a_\lambda-c_\lambda)\cos\varphi_{\bm k}+(b_\lambda+d_\lambda)}{2E_{\bm k}-\omega_\lambda},\label{eq.formalsol1}\\
Y_{\lambda\bm k}=-\frac{g}{2}\frac{(a_\lambda-c_\lambda)\cos\varphi_{\bm k}-(b_\lambda+d_\lambda)}{2E_{\bm k}+\omega_\lambda}.\label{eq.formalsol2}
\end{eqnarray}
We omit $\lambda$ below.
\par
If the p-h symmetric condition (\ref{eq.phcondition}) is satisfied, Eqs.~(\ref{eq.equationX}) and (\ref{eq.equationY}) can be decoupled by introducing the even and odd components as
\begin{eqnarray}
X_{\bm k}^e=(X_{\bm k}+X_{\underline{\bm k}})/2, \quad Y_{\bm k}^e=(Y_{\bm k}+Y_{\underline{\bm k}})/2,\\
X_{\bm k}^o=(X_{\bm k}-X_{\underline{\bm k}})/2,\quad Y_{\bm k}^o=(Y_{\bm k}-Y_{\underline{\bm k}})/2,
\end{eqnarray}
where the former two are even as $X^e_{\underline{\bm k}}=X^e_{\bm k}$ and $Y^e_{\underline{\bm k}}=Y^e_{\bm k}$, while the latter two are odd as $X^o_{\underline{\bm k}}=-X^o_{\bm k}$ and $Y^o_{\underline{\bm k}}=-Y^o_{\bm k}$.
\par
The equations for odd components read
\begin{eqnarray}
&&2E_{\bm k}X_{\bm k}^o-\frac{g}{2}(a-c)\cos\varphi_{\bm k}=\omega X_{\bm k}^o,\label{eq.Xo}\\
&&-2E_{\bm k}Y_{\bm k}^o-\frac{g}{2}(a-c)\cos\varphi_{\bm k}=\omega Y_{\bm k}^o,\label{eq.Yo}\\
&&a=\sum_{\bm k}\cos\varphi_{\bm k}X^o_{\bm k},\quad c=\sum_{\bm k}\cos\varphi_{\bm k}Y^o_{\bm k}.\label{eq.aoco}
\end{eqnarray}
If $a-c\neq 0$, the formal solutions of Eqs.~(\ref{eq.Xo}) and (\ref{eq.Yo}) are given by
\begin{eqnarray}
X_{\bm k}^o=\frac{g}{2}\frac{(a-c)\cos\varphi_{\bm k}}{2E_{\bm k}-\omega},\quad Y_{\bm k}^o=\frac{g}{2}\frac{(c-a)\cos\varphi_{\bm k}}{2E_{\bm k}+\omega}.\label{eq.XYo}
\end{eqnarray}
Setting $X_{\bm k}^e=Y_{\bm k}^e=0$, we obtain $X_{\bm k}=X_{\bm k}^o$ and $Y_{\bm k}=Y_{\bm k}^o$. The condition $b+d=0$, which is obtained from Eqs.~(\ref{eq.formalsol1}) and (\ref{eq.formalsol2}), reduces to 
\begin{eqnarray}
\sum_{\bm k}\frac{\omega\xi_{\bm k}}{E_{\bm k}(4E_{\bm k}^2-\omega^2)}=0.\label{eq.chixy}
\end{eqnarray}
Since Eq.~(\ref{eq.chixy}) is equivalent to $\chi_{xy}(\omega)=0$, it ensures the uncoupled phase and amplitude fluctuations.
\par
Substituting Eq.~(\ref{eq.XYo}) into Eq.~(\ref{eq.aoco}), we obtain
\begin{eqnarray}
1-2g\sum_{\bm k}\frac{\xi_{\bm k}^2}{E_{\bm k}^2}\frac{E_{\bm k}}{4E_{\bm k}^2-\omega^2}=0.
\end{eqnarray}
The above equation is equivalent to $1-2g\chi_{xx}(\omega)=0$ and therefore has the MG mode solution $\omega=2|\Delta_0|$ for which it reduces to the MF gap equation. 
We thus obtain 
\begin{eqnarray}
X_{\bm k}^o&=&-\frac{{A}\cos\varphi_{\bm k}}{2|\Delta_0|-2E_{\bm k}},\quad Y_{\bm k}^o=-\frac{{A}\cos\varphi_{\bm k}}{2|\Delta_0|+2E_{\bm k}},\label{eq.Higgsamplitude}
\end{eqnarray}
where $A$ is the normalization constant. $A$ is determined by the normalization condition (\ref{eq.normalcondition}) as
\begin{eqnarray}
A=\frac{1}{\sqrt{\sum_{\bm k}\frac{|\Delta_0|}{E_{\bm k}\xi_{\bm k}^2}}}.
\end{eqnarray}
The creation operator of the MG mode is thus obtained as
\begin{eqnarray}
\beta_{\rm H}^\dagger=A\sum_{\bm k}\frac{\xi_{\bm k}}{E_{\bm k}}\left(\frac{1}{2|\Delta_0|-2E_{\bm k}}\alpha_{\bm k}^\dagger+\frac{1}{2|\Delta_0|+2E_{\bm k}}\alpha_{\bm k}\right).
\label{eq.NGcreation}
\end{eqnarray}
\par
The equations for even components read
\begin{eqnarray}
&&2E_{\bm k}X_{\bm k}^e-\frac{g}{2}(b+d)=\omega X_{\bm k}^e,\label{eq.Xe}\\
&&-2E_{\bm k}Y_{\bm k}^e+\frac{g}{2}(b+d)=\omega Y_{\bm k}^e,\label{eq.Ye}\\
&&b=\sum_{\bm k}X^e_{\bm k},\quad d=\sum_{\bm k}Y^e_{\bm k}.\label{eq.bede}
\end{eqnarray}
If $b+d\neq 0$, the formal solutions of Eqs.~(\ref{eq.Xe}) and (\ref{eq.Ye}) are given by
\begin{eqnarray}
X_{\bm k}^e=\frac{g}{2}\frac{b+d}{2E_{\bm k}-\omega},\quad Y_{\bm k}^e=\frac{g}{2}\frac{b+d}{2E_{\bm k}+\omega}.\label{eq.formalsole}
\end{eqnarray}
Setting $X_{\bm k}^o=Y_{\bm k}^o=0$, we obtain $X_{\bm k}=X_{\bm k}^e$ and $Y_{\bm k}=Y_{\bm k}^e$. The condition $a-c=0$, which is obtained from Eqs.~(\ref{eq.formalsol1}) and (\ref{eq.formalsol2}), reduces to Eq.~(\ref{eq.chixy}).
\par
Substituting Eq.~(\ref{eq.formalsole}) into Eq.~(\ref{eq.bede}), we obtain
\begin{eqnarray}
1-2g\sum_{\bm k}\frac{E_{\bm k}}{4E_{\bm k}^2-\omega^2}=0.
\end{eqnarray}
The above equation is equivalent to $1-2g\chi_{yy}(\omega)=0$ and therefore has the NG mode solution $\omega=0$, for which it reduces to the MF gap equation. We thus obtain
\begin{equation}
X_{\bm k}^e=Y_{\bm k}^e=A'/E_{\bm k},\label{eq.NGamplitude}
\end{equation}
where $A'$ is the normalization constant. However, Eq.~(\ref{eq.NGamplitude}) does not fulfill the normalization condition (\ref{eq.normalcondition}). This anomaly is typical for zero energy modes. It can be avoided by introducing a small fictitious external field in the Hamiltonian (\ref{eq.HBCS_spin}) \cite{zeromode}. The creation operator of the NG mode is thus obtained as
\begin{eqnarray}
\beta_{\rm NG}^\dagger=A'\sum_{\bm k}\frac{1}{E_{\bm k}}(\alpha_{\bm k}^\dagger+\alpha_{\bm k}).
\label{eq.NGcreation}
\end{eqnarray}
\par
In the limit of small fluctuation $\alpha_{\bm k}^\dagger\alpha_{\bm k}\ll 1$, using $\alpha_{\bm k}^\dagger\simeq S'^+_{\bm k}$ and $\alpha_{\bm k}\simeq S'^-_{\bm k}$, the creation operators for the Higgs mode and the NG mode can be obtained as Eqs.~(\ref{eq.betaH}) and (\ref{eq.betaNG}).


\begin{thebibliography}{}
\bibitem{higgsdef}
In this paper, a gapful collective mode that dominantly induces amplitude oscillation of an order parameter is referred to as a massive Goldstone (MG) mode. It may induce both amplitude and phase oscillation of a complex order parameter. Specifically, when it induces only amplitude oscillation, we call it a pure amplitude MG mode.
\bibitem{fullH}
We remark that the full Hamiltonian
$\tilde{\mathcal H}=\sum_{\bm k,\sigma}\xi_{\bm k}c_{\bm k\sigma}^\dagger
 c_{\bm k\sigma} -g \sum_{\bm k,\bm k',\bm q} c_{\bm k+\bm q\upa}^\dagger c_{-\bm
 k+\bm q\dna}^\dagger c_{-\bm k'+\bm q\dna}c_{\bm k'+\bm q\upa}$,
is invariant by all the discrete transformations $\mathcal C$, $\mathcal P$, $\mathcal T$, and $\Theta$. The extension of the argument in this paper to more general cases for $\tilde{\mathcal H}$ is straightforward.
\bibitem{Pvac}
For the BCS wave function $|\Psi\rangle$, $\mathcal P|\Psi\rangle=|\bar\Psi\rangle$ holds if $\prod_{\bm k}(-1)=1$.
\bibitem{cooperon}
$|\phi_{\rm H}\rangle$ and $|\phi_{\rm NG}\rangle$ form the bound states inside the band gap, which are referred to as ``Cooperons", if the fermionic dispersion has a band gap or Dirac cones. Evolution of Cooperons into the MG and NG modes in a Dirac fermion system across the quantum critical point was studied in Ref.~\onlinecite{tsuchiya-13}.
\bibitem{degeneracy}
$1-2g\chi_{xx}(\omega)=0$ and $1-2g\chi_{yy}(\omega)=0$ in the normal phase ($\Delta_0=0$) give the poles that correspond to the excitation energy of $|\phi_{\rm H}\rangle$ and $|\phi_{\rm NG}\rangle$, respectively. $\chi_{xx}(\omega)=\chi_{yy}(\omega)$ in the normal phase due to the U(1) symmetry therefore indicates the degeneracy of $|\phi_{\rm H}\rangle$ and $|\phi_{\rm NG}\rangle$. In fact, within the MF theory, Eq.~(\ref{eq.suscept_xxyy}) leads to $\chi_{xx}(\omega)=\chi_{yy}(\omega)=\sum_{\bm k}|\xi_{\bm k}|/(4\xi_{\bm k}^2-\omega^2)$.
\bibitem{CPTcommutation}
Note that $[\mathcal C,\mathcal T]= [\mathcal C,\mathcal P]=[\mathcal T,\mathcal P]=0$ and simultaneous eigenstates of $\mathcal C$, $\mathcal T$, and $\mathcal P$ exist if $\prod_{\bm k}(-1)=1$.
\end{thebibliography}
\end{document}